\documentclass[notitlepage,twocolumn,prl,tightenlines,nofootinbib,superscriptaddress]{revtex4}

\usepackage{amsmath}
\usepackage{amssymb,amsfonts}
\usepackage{bm}
\usepackage[mathcal]{euscript}
\usepackage{graphicx}
\usepackage{subfigure}
\usepackage{hyperref}

\newcommand{\eg}{\textit{e.g.}}

\newcommand{\mathnotation}[2]{\newcommand{#1}{\ensuremath{#2}}}

\mathnotation{\ldef}{\mathrel{\raisebox{.069ex}{:}\!\!=}}
\mathnotation{\rdef}{\mathrel{=\!\!\raisebox{.069ex}{:}}}
\mathnotation{\id}{\mathrm{identity}}		
\mathnotation{\bi}{b}				
\mathnotation{\wb}{w_{\mathrm{B}}}		
\mathnotation{\lm}{\lambda_{M}}			
\mathnotation{\tw}{t_{\mathrm{white}}}          
\def\d{{\mathrm d}}

\begin{document}

\title{Walls inhibit chaotic mixing}

\author{E. Gouillart}
\author{N. Kuncio}
\author{O. Dauchot}
\author{B. Dubrulle}
\affiliation{Service de Physique de l'Etat Condens\'e, DSM, CEA Saclay, URA2464, 91191
Gif-sur-Yvette Cedex, France}
\author{S. Roux}
\affiliation{Surface du Verre et Interfaces, UMR CNRS/Saint-Gobain, 93303 Aubervilliers, France}
\author{J.-L. Thiffeault}
\affiliation{Department of Mathematics, Imperial College London, SW7 2AZ,
United Kingdom}

\date{\today}


\begin{abstract}

  We report on experiments of chaotic mixing in a closed vessel, in
  which a highly viscous fluid is stirred by a moving rod. We analyze
  quantitatively how the concentration field of a low-diffusivity dye
  relaxes towards homogeneity, and observe a slow algebraic decay of
  the inhomogeneity, at odds with the exponential decay predicted by
  most previous studies. Visual observations reveal the dominant role
  of the vessel wall, which strongly influences the concentration
  field in the entire domain and causes the anomalous scaling. A
  simplified 1-D model supports our experimental results. Quantitative
  analysis of the concentration pattern leads to scalings for the
  distributions and the variance of the concentration field consistent
  with experimental and numerical results.

\end{abstract}

\maketitle

Low-Reynolds-number fluid mixing has a variety of applications ranging
from geophysics to industrial mixing devices. While turbulent flows lead
to highly efficient mixing, simple laminar flows with chaotic Lagrangian
dynamics also promote rapid homogenization \cite{Aref1984}. Dynamical
systems approaches based on flow kinematics have provided a first insight
into chaotic mixing \cite{Ottino1989, Jana1994}.  A deeper understanding of
homogenization processes is gained by examining the interplay between
chaotic stirring and diffusion. Several experimental \cite{Rothstein1999,
Jullien2000} and numerical \cite{Fereday2002, Fereday2004} studies
obtained an exponential decay for the variance of a diffusive scalar
concentration field in a chaotic mixer. This behavior is attributed to an
 asymptotic spatial structure of the scalar dubbed \emph{strange eigenmode}
\cite{Pierrehumbert1994}, that results in a global exponential decay of
the spatial contrast.  However, these theories focus on ideal mixing
systems, \eg~with periodic or slip boundary conditions, far from the reality of
industrial mixing devices with solid no-slip walls. It has been suggested
\cite{Chertkov2003a, Schekochihin2004} that mixing might be slower in
bounded flows, but experimental evidence is still lacking.

In this Letter, we study experimentally dye homogenization by chaotic
mixing in a 2-D closed flow. Precise measurements of the
concentration field yield ``slow'' algebraic decay of an
inhomogeneity, at odds with the expected exponential decay. We relate
quantitatively this slow mixing to the chaotic nature of trajectories
initially close to the no-slip wall, which end up escaping in the bulk
and slow down the whole mixing process.

\begin{figure}
\includegraphics[width=0.46\columnwidth]{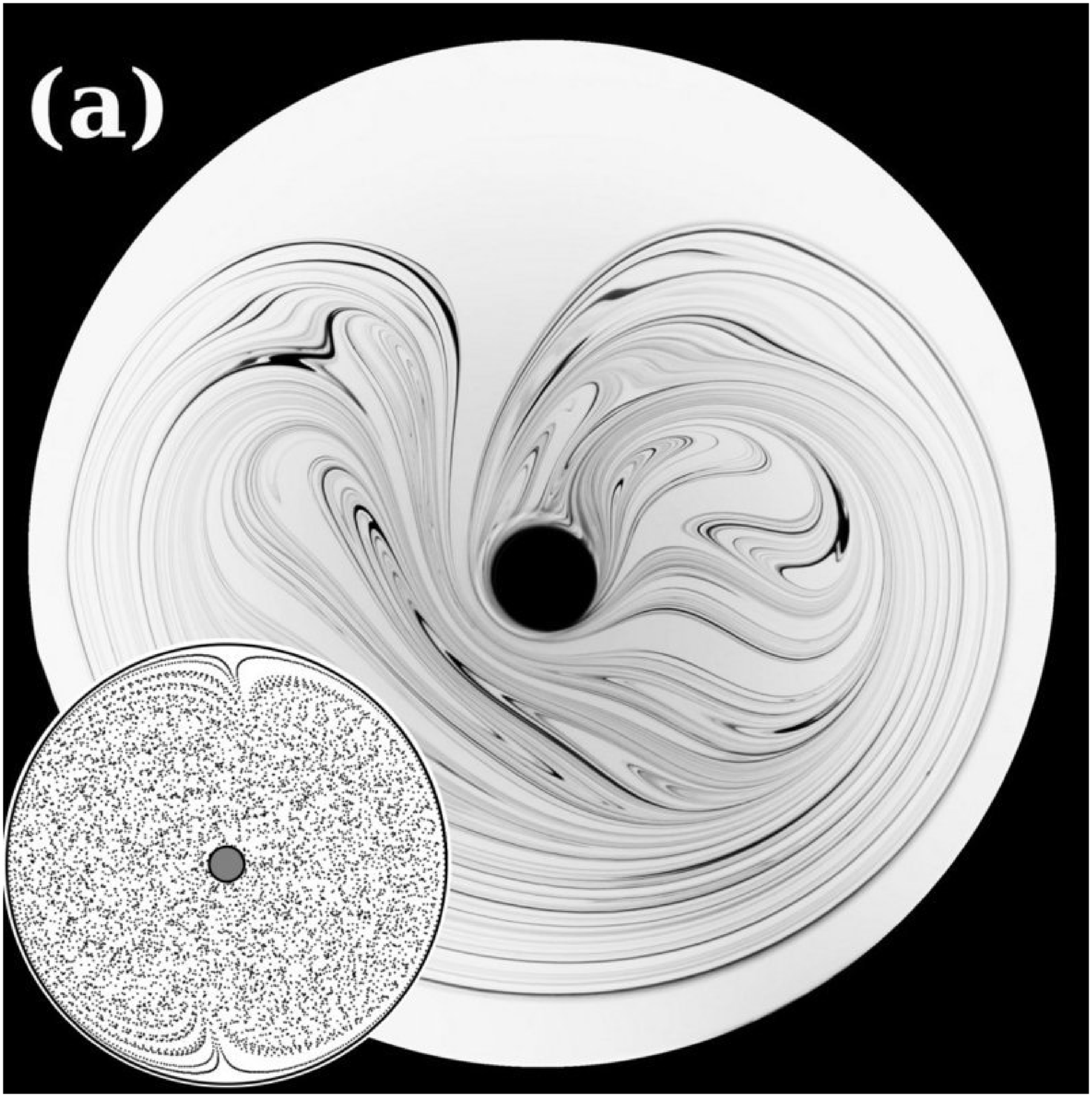}
~\includegraphics[width=0.46\columnwidth]{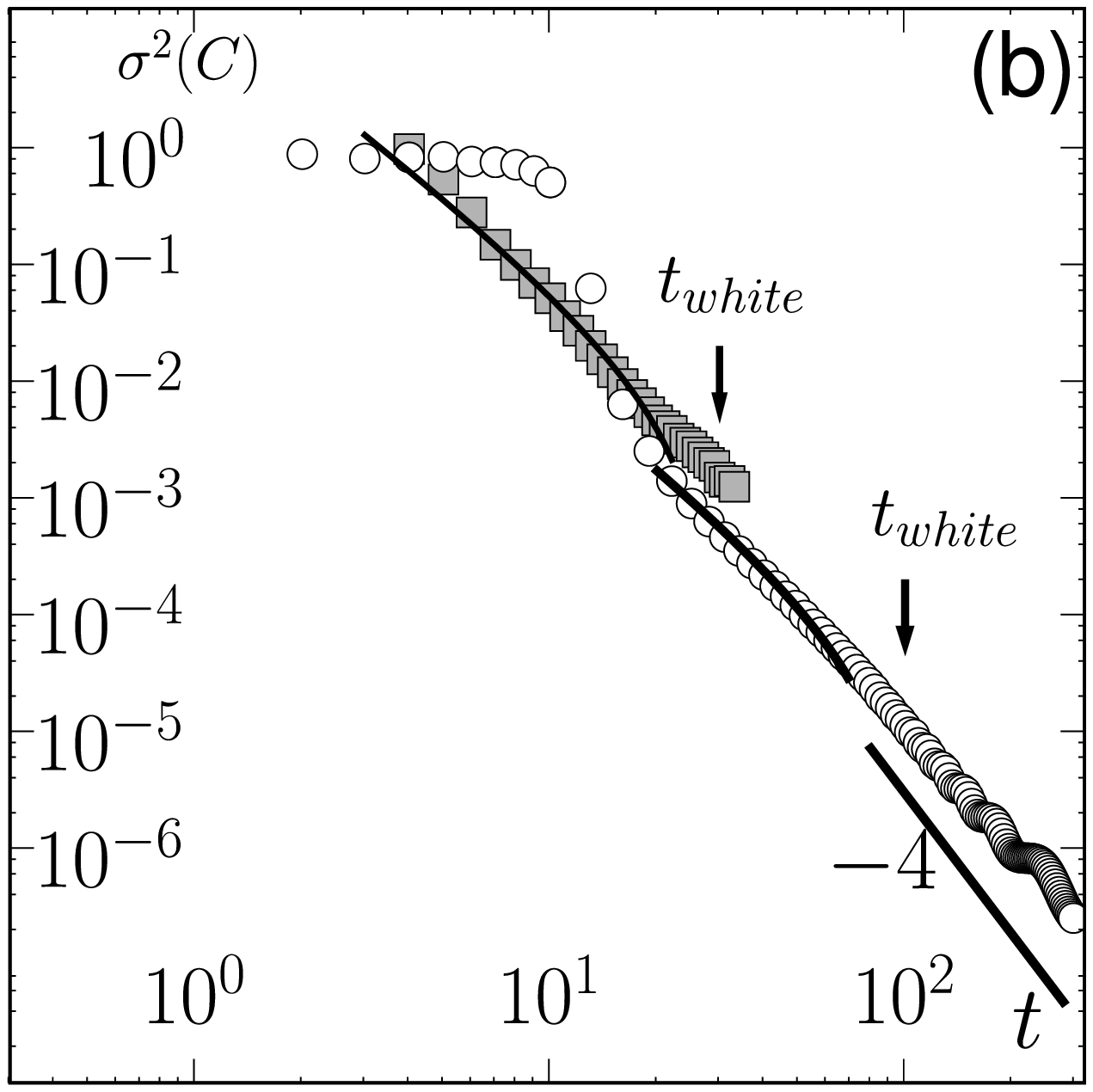}
\includegraphics[width=0.99\columnwidth]{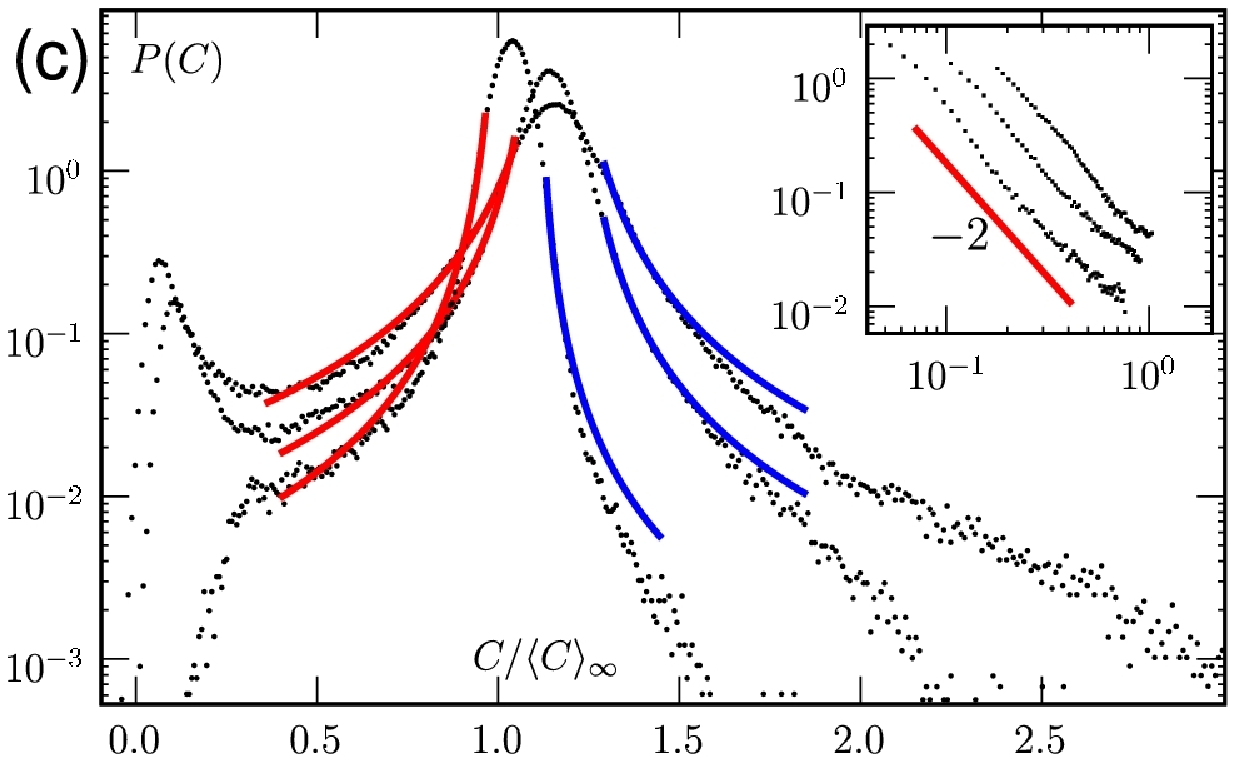}
\caption{(a) Chaotic mixing experiment in a closed vessel: a rod moves
periodically on a figure-eight path (see Fig. \ref{fig:prof} (a)) and
transforms an initial spot of dye into a complicated filamentary pattern.
Inset: Poincar\'e section obtained numerically for the corresponding
Stokes flow. Note that the chaotic region spans the entire domain. 
(b) Evolution of the variance of the concentration field in a fixed central region.
{\tiny $\blacksquare$}: experiment
, {\small $\circ$}: numerical simulation (see Fig.~\ref{fig:PDF_model} for a
description). Solid line fits: contribution of the white pixels
$\sigma^2_W\simeq (2\log{t}+\log{\wb})/(\log{\Gamma}\times t^2)$ derived
below.
(c) Experimental concentration PDFs in the bulk  at time $t=13,17,31$
periods. Both sides of the peak can be fitted by power laws
$(C_{max}-C)^{-2}$ (red and blue plots). Inset: left (``light-gray'')
tail of the peak, $P(C)$ against $|C_{max}(t)-C|$.\label{fig:mixer}}
\vspace{-0.3cm}
\end{figure}

A cylindrical rod periodically driven on a figure-eight path gently
stirs viscous sugar syrup inside a closed vessel (Fig.~\ref{fig:mixer}
(a)).  The stirring scale is comparable to the vessel size, in
contrast to other devices such as arrays of magnets
\cite{Rothstein1999, Jullien2000}. This protocol is a good candidate
for efficient mixing: we can observe on a Poincar\'e section
\cite{Gouillart2006} (Fig.~\ref{fig:mixer} (a)) -- computed
numerically for the corresponding Stokes flow -- a large chaotic
region spanning the \emph{entire} domain, including the vicinity of
the wall. The signature of chaotic advection can also be observed in
Fig.~\ref{fig:mixer} (a), where a complex lamellar pattern is created
by the stretching and folding of an initial dye blob into
exponentially thin filaments. The fluid viscosity $\nu=5\times
10^{-4}\, \text{m}^2\cdot\text{s}^{-1}$ together with rod diameter
$\ell=16 \,\text{mm}$ and stirring velocity $U=2~
\text{cm}\cdot\text{s}^{-1}$ yield a Reynolds number $Re = U\ell/\nu
\simeq 0.6$, consistent with a Stokes flow regime. A spot of
low-diffusivity dye (Indian ink diluted in sugar syrup) is injected at
the surface of the fluid, and we follow the evolution of the dye
concentration field during the mixing process (see
Fig.~\ref{fig:mixer}).  The concentration field is measured through
the transparent bottom of the vessel using a $12$-bit CCD camera at
resolution $2000 \times 2000$.

 Despite the exponential stretching occurring in the
bulk, the resulting variance $\sigma^2(C)$ of the concentration field
(measured in a large central rectangular region) decays surprisingly
slowly with time $t$ as a power law $t^{-m}$ with $m\simeq 3.2$
(Fig.~\ref{fig:mixer} (b)), and not exponentially as expected. This behavior persists until the end of the experiment (35
periods), by which time the variance has decayed by more than three orders
of magnitude. Moreover, concentration probability distribution functions
(PDFs) shown on Fig.~\ref{fig:mixer} (c) exhibit wide power-law tails on
both sides of the most probable value. The probability
of ``white'' (zero) concentration decays very slowly with time, whereas
the peak shifts towards lower values.

\begin{figure}
\begin{minipage}{0.5\columnwidth}
\includegraphics[width=0.85\textwidth]{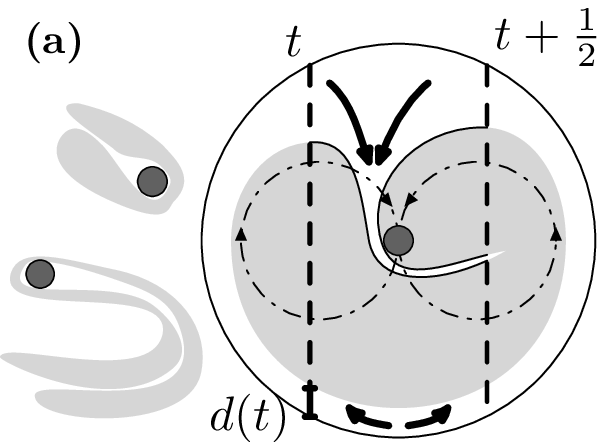}
\end{minipage}
\begin{minipage}{0.47\columnwidth}
\includegraphics[width=0.90\columnwidth]{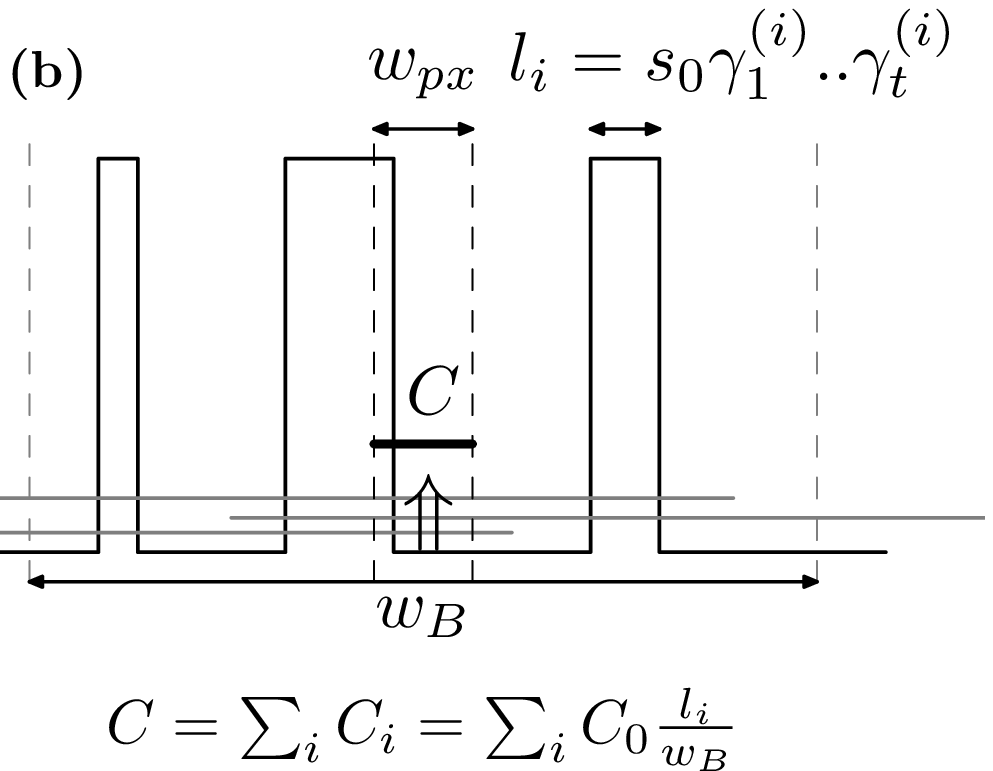}
\end{minipage}
\caption{(a) Transport mechanisms: (i) the rod stretches and folds the
mixed region and (ii) a white strip of unmixed fluid is injected between
the two folded parts. (b) Stretched strips of dye (black) are smeared
out by diffusion on a scale $\wb$ (gray). The concentration $C$ of a
pixel $x$ is then given by
adding the concentrations coming from strips inside a box of
size $\wb$ around $x$.  
\label{fig:prof}}
\vspace{-0.3cm}
\end{figure}

In order to understand these surprising scalings, we first describe the
various mechanisms at play during the mixing process. A blob of dye,
initially released close to the vessel center, is transformed into a
complicated pattern expanding towards the wall with time. We
distinguish at each instant the growing ``mixed region'', delimited by
the advected blob frontier, and the remaining wall region where
$C=0$, in the vicinity of the vessel wall. This distinction is obvious in
Fig.~\ref{fig:mixer} (a) where one can observe a central heart-shaped
mixed region and an annular unmixed wall region. As the chaotic
region spans the entire flow, fluid particles initially close to the
zero-velocity wall eventually escape from the wall region to wander through
the whole chaotic region. Trajectories escape along the unstable manifold
of parabolic separation points on the wall \cite{Jana1994,
Haller2004}. The signature of such an escape path can be visualized on
Fig.~\ref{fig:mixer} (a) where unmixed fluid from the wall region is
``sucked'' inside the mixed region through its white cusp, close to the
rod. This results in the periodic injection of broad white strips that
can be observed inside the mixed region. The mixed region then grows
towards the wall to make up for this mass injection.
Incompressibility combined with zero-velocity condition at the wall
leads \cite{Chertkov2003a} to a shrinking distance between the mixed
region border and the wall scaling as $d(t)\propto t^{-1}$. This
scaling is verified experimentally. The area of unmixed fluid from the
wall region injected at each period inside the mixed region then scales as
$\dot{d}(t) \propto t^{-2}$. As a result, the mean concentration value
inside the bulk decreases with time as $(1-d(t))^{-1}$. Simultaneously,
the mixed region is stretched and folded at each half-cycle of the rod
movement (see Fig.~\ref{fig:prof} (a)) in a baker's-map-like fashion
\cite{Farmer1983}. However, it should be noted that the two folded parts
are not stacked directly onto each other but \emph{separated by the
newest injected white strip}. Also note that the part ``attached'' to the
rod has experienced more stretching than the one ``left behind''.
Briefly, (i) chaotic stretching imposes that the typical width of a dye
filament in the bulk shrinks exponentially down to the diffusion or
measurement scale, yet (ii) wide strips of unmixed fluid of width
$\dot{d}(t)\propto 1/t^2$ are periodically inserted between these fine
structures. In the following we derive how these two effects lead to the
observed scalings.

\begin{figure}
\includegraphics[width=0.99\columnwidth]{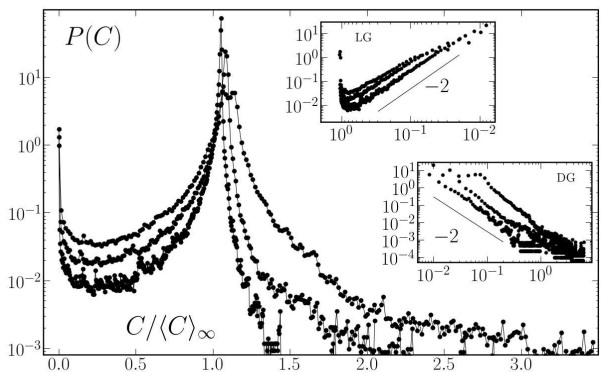}
\caption{\label{fig:PDF_model} Numerical concentration PDFs
($t=15,\,18,\,25$) obtained by
letting concentration profiles evolve as $C(x,t+1)=C(f^{-1}(x),t)$, 
$f(x):
f_1(x)=  x - ax^2 + (\gamma-1 +a)x^3 ;
f_2(x)=  1 - ax^2 + (\gamma-1 +a)x^3 $ with $a=0.9$ and
$\gamma=0.55$. Upper (resp. lower) inset: light (resp. dark) gray tail, $P(C)$ against
$|C_{max}-C|$. We observe the same power-law decay $(C_{max}-C)^{-2}$ on
both sides of the maximum as in the experiment (Fig.~\ref{fig:mixer}
(c)).
}
\vspace{-0.3cm}
\end{figure}

For this purpose, we simplify the 2-D problem by characterizing only 1-D
concentration profiles $C(x,t)$ along a secant to the stretching
direction -- the dashed segments on Fig.~\ref{fig:prof} (a) -- that is,
we neglect the variation of the concentration along a dye filament on a
scale comparable to the vessel size. We thus call from now on  ``mixed
region'' the intersection of the 2-D mixed region with such a segment.
The effect of the mixer during a half-period then amounts to the action
of a one-dimensional discrete-time map that transforms concentration
profiles by inserting an interval of width $\dot{d}_t$ of fluid from the
wall region between two inhomogeneously compressed images of the mixing
region at the previous time (see Fig. \ref{fig:prof} (a)). Such a map
$f$, defined on $[0,1]$ for simplicity, evolves concentration profiles as
$C(x,t+1)=C(f^{-1}(x),t)$ and meets the following requirements: (i) it
is a continuous double-valued function to account for the
stretching/folding process; (ii) $x=0$ is a marginally unstable point of
$f^{-1}$; the correct dynamics close to the wall are indeed
reproduced by imposing $f^{-1}(x)\simeq x + ax^2 + \cdots,\; a>0$ for
small $x$; (iii) because of mass conservation, at each $x$, the local
slopes of the two branches add up to $1$. Other details of $f$ are
unessential for our discussion. Diffusion is mimicked by letting the
concentration profile diffuse between successive iterations of the map.
This model is a modified baker's map
\cite{Farmer1983}, with a parabolic point at $x=0$, whereas the
dynamics are purely hyperbolic in a classical baker's map. Numerical
simulations for a specific choice of $f$ lead to results shown in
Fig.~\ref{fig:mixer} (b) and Fig.~\ref{fig:PDF_model}: both the power-law
evolution of the variance and the different aforementioned features of
the experimental PDFs are reproduced by the simulations.


The map transforms an initial blob of dye of width $s_0$ into an
increasing number of strips with widths $s_0 \gamma_1 \cdots
\gamma_t$, resulting from different
stretching histories inside the mixed region, where $\gamma_t$ is the
compression experienced at time $t$. White
strips also experience this multiplicative stretching from their
injection time. As the mixed region grows towards the wall,
different values of stretching will be sampled. It will nevertheless be
justified below that the concentration measures only require knowledge of the
stretching histories traced back during a finite number of periods. This
allows us to define in a ``quasi-static'' approximation a slowly varying
instantaneous ``Lyapunov exponent'' $\Gamma(t)=\exp(\langle -
\log(|\frac{\partial f^{-1}(x)}{\partial x}|)\rangle_{MR}) $, that is the
geometric mean of the compression taken over the mixed region (MR) at
time $t$. Note that as the white strips are injected close to the center
of the domain, the two branches of $f$ have comparable mean slopes,
yielding the estimate $\Gamma \sim 0.5$. Without diffusion, dye strips
would have a typical width $s_0 \Gamma^t$ at time $t$. However, the
balance between stretching and diffusion imposes that the width of a
strip stabilizes at the Batchelor scale $\wb =
\sqrt{\kappa/(1-\Gamma^2)}$, where $\kappa$ is the diffusion coefficient.
$\wb$ is thus the smallest lengthscale that can be observed in the
concentration profile, and different elementary strips may overlap
(Fig.~\ref{fig:prof} (b)). Since the concentration field is probed by
averaging it on the pixel size $w_{px}$, which is smaller than $\wb$, the
concentration at a pixel is given by adding the contributions from strips
contained in a box of size $\wb$ around the pixel. Hence we characterize
$P(C)$ by considering the different combinations of strip widths for a
zero-diffusivity dye -- the concentration profile on Fig.~\ref{fig:prof}
(b) -- that one might find in a box of size $\wb$. We will distinguish
between three generic cases corresponding to three different regions of
the histogram $P(C)$ (see Fig.~\ref{fig:mixer} (c) and
Fig.~\ref{fig:PDF_model}): a white (W) peak at $C=0$ corresponding to
injected white strips still wider than $\wb$, light gray (LG) and dark
gray (DG) tails corresponding to respectively smaller and larger
concentrations than the peak (mean) concentration. Once we have
quantified the proportion of boxes contributing to these
different values of $C$, the variance will be readily obtained as
\begin{equation} 
\sigma^2(C)=\int (C-\langle C \rangle)^2 P(C) \d C=
\sigma^2_W + \sigma^2_{LG} + \sigma^2_{DG}. \label{eq:var}
\end{equation}

Let us start with white (zero) concentration measures that come from the
stretched images of white strips injected before $t$. White strips
injected at an early time have been stretched and wiped out by diffusion,
that is their width has become smaller than $\wb$. Hence the oldest white
strips that can be observed have been injected at the time $t_i(t)$ such
that $\dot{d}_{t_i}\Gamma^{t-t_i}=\wb$. Note that from $\tw$ defined by
$\dot{d}_{\tw}=\wb$, the injected white strip is smaller than $\wb$ and
no white pixels can be observed. Before $\tw$, the number of white pixels
is proportional to $n_W = d_{t_{i}(t)} - d_{t}\propto(t -t_{i})(t_{i}
t)^{-1}$ for large $t$ (using $d_t \propto t^{-1}$). As
$t-t_i \simeq (2\log{t}+\log{\wb})/\log{\Gamma}$, $n_W \simeq
(2\log{t}+\log{\wb})/(\log{\Gamma}\times t^2)$. We deduce
$\sigma^2_{W}\simeq (2\log{t}+\log{\wb})/(\log{\Gamma}\times t^2)$ for $t<\tw$ and
$\sigma^2_{W}=0$ after $\tw$.  

We now concentrate on the distribution of light gray values
corresponding to white strips that have just been compressed below the
cut-off scale $\wb$.  We propose to approximate the measured value $C$ as the
average of the biggest white strip with width $\lambda < \wb$, and
mixed "gray" fluid whose concentration is close to the most probable
concentration $C_{\mathrm{g}}$. A box with a white strip of scale $\lambda$ thus
bears a
concentration $C_{\lambda}=C_{\mathrm{g}}(1-\lambda/\wb)$ and we can relate
$P(C)$ to the distribution of widths of the images of the injected white
strips $Q(\lambda)$. A white strip injected at $t_0$ is transformed into
$2^{t-t_0}$ images with scales $\dot{d}_{t_0} \Gamma^{t-t_0}$.
Therefore
$
Q(\lambda)=(\lambda/\dot{d}_{t_{0}})^{\log(2)/\log(\Gamma)}
\times (1/\lambda \log{\Gamma})
$,
resulting in
\begin{eqnarray}
P(C) &=& \dot{d}_t^{\,\log{2}/\log{(\Gamma^{-1})}}\frac{\wb}{C_{\mathrm{g}}(t)}
\bigg[\wb (1 - \frac{C}{C_{\mathrm{g}}}) \bigg
]^{\frac{\log{2}}{\log{\Gamma}}-1}\nonumber \\
&=& g(t) \bigg [C_{\mathrm{g}} - C \bigg
]^{(\log{2}/\log{\Gamma})-1}\;.
\label{eq:PC}
\end{eqnarray}
$P(C)$ thus has a power-law tail in the light gray levels whose exponent
depends on the mean stretching $\Gamma$. We observe satisfactory
agreement between this prediction and both experimental data and
numerical 1-D simulations (see Fig.~\ref{fig:mixer} (c) and
Fig.~\ref{fig:PDF_model}) where for this tail $P(C)\propto (C-C_{\mathrm{g}})^{-\alpha}$ with
$\alpha \lesssim 2$,
consistent with $\Gamma \lesssim 0.5$, a rather homogeneous
stretching. Also note that the amplitude of the light-gray tail decreases
as  a power law $g(t)\propto \dot{d}_t^{\,\log 2/\log(1/\Gamma)} \propto t^{-2(\log
2/\log(1/\Gamma))}$. We deduce 
\[
\sigma^2_{LG}=g(t)\int_{C_{\mathrm{min}}}^{C_{\mathrm{g}}} (C_{\mathrm{g}}-C)^{2-\alpha(\Gamma)} \d C
\]
where $\alpha(\Gamma)=1-\log{2}/\log{\Gamma}$, and $C_{\mathrm{min}}$ is the
smallest concentration observed ($C_{\mathrm{min}}=0$ for
$t<\tw$ and $C_{\mathrm{min}}=C_{\mathrm{g}}(1-\dot{d}_t/\wb)$ for $t>\tw$). For $t<\tw$ the
integral is constant and $\sigma^2_{LG}\propto g(t)
\propto \dot{d}_{t} \propto t^{-2}$. On the other hand, for $t\ge\tw$,
\[
\sigma^2_{LG}=\frac{g(t)}{2+\alpha(\Gamma)}[C_{\mathrm{g}}-C_{\mathrm{min}}]^{\,3-\alpha(\Gamma)}
        \propto t^{-(6+2\frac{\log 2}{\log \Gamma})}\,.
\]
For $\alpha(\Gamma)\sim 2$ as we observed, the exponent in the above power law
is about $-4$.

Finally, the dark gray tail corresponds to boxes 
containing images of mixed regions from early times -- thus with an
important percentage of black --
that have experienced little stretching.
It is therefore not sufficient
to consider only the mean stretching $\Gamma$ as before, since stretching histories
far from the mean are involved. Using the large-deviation
function $S$ for the 
finite-time Lyapunov exponents distribution \cite{Ott1994}, we derive
\[
P(C,t)=\frac{2^{t-t_0}\exp[-(t-t_0)S(-\frac{\log \wb}{t-t_0}
+\log \Gamma)]}
{(C-\langle C \rangle)^{2}(t-t_0)^{2}},
\]
where $t_0(C)$ is the earlier time at which the mixed region had a mean
concentration $C$, so that $C=\langle C \rangle (1-d_{t_0})^{-1}$.
At a
fixed time the dark gray tail decreases as the dominant contribution
$(C-\langle C \rangle)^{-2}$, however the probability to observe a fixed
concentration value decays exponentially, allowing
us to neglect $\sigma^2_{DG}$ in (\ref{eq:var}). Both the
$(C-\langle C \rangle)^{-2}$ shape and the rapid fall-off of the
dark-gray tail can be observed on Fig.~\ref{fig:mixer} (c) and
Fig.~\ref{fig:PDF_model}.

We now sum these contributions to obtain $\sigma^2(C)$.
In the experiment, the crossover 
$\tw$ is estimated as $30$ periods. However, 3-D effects inside the fluid
prevented us from conducting experiments for more than $35$ periods.
For this early regime, fitting the data with $\sigma^2_{W}\propto
(2\log{t}+\log{\wb})/t^2$ (black line on
Fig.~\ref{fig:mixer} (b)) gives good results,
except close to $\tw$ where the contribution of the
light gray tail starts to dominate.
In contrast, in numerical simulations
we observe (Fig.~\ref{fig:mixer} (b)) both the
$(2\log{t}+\log{\wb})/t^2$ behavior (black line) , which can be
interpreted as in the experiment, and the $t^{-4}$ decay after
$\tw$ ($100$ periods for the case studied) given by $\sigma^2_{LG}$.  

Note finally that the algebraic nature of $d(t)$ permits crude ``first-order''
approximations such as considering only the mean stretching given by the
Lyapunov exponent as we did. The injection process dominates other mechanisms
put forward to analyze concentration distributions, such as the evolution
of the distribution by self-convolution due to the random addition of
concentration levels \cite{Villermaux2003, Venaille2006}. The strange
eigenmode formalism also fails to describe this nonasymptotic regime as
the spatial mixing pattern is still evolving. 

In conclusion, we propose a scenario for 2-D mixers with a chaotic region that extends to fixed walls. As soon as the flow is sufficiently slow close to the wall -- a rather generic situation -- and keeps for long times an unmixed pool close to a parabolic point, the unstable manifold of which feeds fluid into the mixing region, the entire concentration field is affected, regardless of distance from the walls. The algebraic scalings for the variance and concentration distributions can be predicted from the filamentary stirring pattern generated by a combination of stretching, folding, and injection of fluid from the fixed walls.
Our reasoning could be extended to other algebraic expressions of
$d(t)$ resulting from different hydrodynamics at the wall. In the
present case (no-slip wall), we derive a power-law (asymptotically
$t^{-4}$) evolution for the decay of the concentration variance, and
find very good agreement between our analytical, experimental, and
numerical results.  Our simulations of the well-studied viscous
blinking vortex flow \cite{Jana1994} also yield the same algebraic
decay for the variance of a coarse-grained concentration, obtained by
advecting a large number of points. These numerical results will be
presented in a future paper, where we will generalize our approach to
other systems characterized by continuous injection of inhomogeneity,
such as open flows.

The authors thank E. Villermaux and F.  Daviaud for fruitful
discussions.

\end{document}